\PassOptionsToPackage{dvips}{graphicx}
\documentclass{ifacconf}
\usepackage{amsmath,amssymb,amsfonts,mathrsfs,mathtools}
\usepackage{url}
\makeatletter
\let\old@ssect\@ssect 
\makeatother
\usepackage{refcount}
\usepackage{natbib}        
\usepackage[dvipdfmx,colorlinks=true,linkcolor=blue,citecolor=blue,urlcolor=blue]{hyperref}
\makeatletter
\def\@ssect#1#2#3#4#5#6{%
  \NR@gettitle{#6}
  \old@ssect{#1}{#2}{#3}{#4}{#5}{#6}
}
\makeatother
\usepackage{muracommands}
\begin{document}
\begin{frontmatter}

\title{The Waterbed Effect on Quasiperiodic Disturbance Observer: Avoidance of Sensitivity Tradeoff with Time Delays\thanksref{footnoteinfo}}

\thanks[footnoteinfo]{This work was supported by JSPS KAKENHI Grant Number JP25K17560.\\
{\color{red}
  This paper has been published in IFAC Journal of Systems and Control, vol. 35, p. 100392, Mar. 2026.\\\indent
  See https://doi.org/10.1016/j.ifacsc.2026.100392
}}

\author[First]{Hisayoshi Muramatsu}
\address[First]{Mechanical Engineering Program, Hiroshima University,
Higashihiroshima, Hiroshima, 739-8527, Japan (e-mail: muramatsu@hiroshima-u.ac.jp).}
\begin{abstract} 
In linear time-invariant systems, the sensitivity function to disturbances is designed under a sensitivity tradeoff known as the waterbed effect.
To compensate for a quasiperiodic disturbance, a quasiperiodic disturbance observer using time delays was proposed.
Its sensitivity function avoids the sensitivity tradeoff, achieving wideband harmonic suppression without amplifying aperiodic disturbances or shifting harmonic suppression frequencies.
However, its open-loop transfer function is not rational and does not satisfy the assumptions of existing Bode sensitivity integrals due to its time delays.
This paper provides Bode-like sensitivity integrals for the quasiperiodic disturbance observer in both continuous-time and discrete-time representations and clarifies the avoided sensitivity tradeoff with time delays.
\end{abstract}


\end{frontmatter}
\section{Introduction}
One of the fundamental issues in automatic control is disturbance compensation.
In practical systems, disturbances always exist and impair the precision of automatic control.
For linear time-invariant systems, we design a sensitivity function $S(s)\ceq1/(1+\Gamma(s))$ based on an open-loop transfer function $\Gamma(s)$, while tradeoffs restrict the design.
One tradeoff lies between the sensitivity function $S(s)\ceq1/(1+\Gamma(s))$ and complementary sensitivity function $T(s)\ceq\Gamma(s)/(1+\Gamma(s))$, which are restricted by the unit sum $S(s)+T(s)=1$.
The sensitivity and complementary sensitivity functions represent disturbance-suppression performance and robust stability against modeling errors, respectively.
It is necessary to reduce their gains appropriately to improve performance under the unit-sum tradeoff.
For the sensitivity function, there exists another tradeoff in the sensitivity integral given by \cite{bode1945network}.
This Bode sensitivity integral (\cite{Freudenberg1985,DING202297,NAGARSHETH2021100036}) states that a stable sensitivity function satisfies
\begin{align}
	\label{eq:Bode:integral:conv:CT}
	\int_{0}^{\infty} \ln |S(j\omega)| d\omega
	= \pi \sum_{i} \rmRe[q_i] - \frac{\pi}{2}\lim_{s\to\infty}s\Gamma(s),
\end{align}
where $q_i$ are the unstable right-half-plane poles of a proper rational function $\Gamma(s)$.
Also, \cite{wu1992simplified,SUNG01121988,mohtadi1990bode,Emami2019} provide the Bode sensitivity integral for discrete-time control.
Based on a rational open-loop transfer function $\tilde{\Gamma}(z)$, the stable sensitivity function $\tilde{S}(z)\ceq1/(1+\tilde{\Gamma}(z))$ satisfies
\begin{align}
	\label{eq:Bode:integral:conv:DT}
	\int_{0}^{2\pi} \ln |\tilde{S}(e^{j\omega})| d\omega
	&= 2\pi \sum_{i} \ln|\tilde{q}_i|\notag\\
	&\hspace{1em} - 2\pi \ln|\lim_{z\to\infty}\tilde{\Gamma}(z)+1|,
\end{align}
where $\tilde{q}_i$ are unstable poles outside the closed-unit disk of a proper rational function $\tilde{\Gamma}(z)$.
The Bode sensitivity integral is called the waterbed effect (\cite{skogestad2005multivariable,COSTACASTELLO2015259}), as the gain increases at some frequencies while decreasing at other frequencies.
However, these Bode sensitivity integrals are applicable only to proper rational open-loop transfer functions.
Periodic disturbances often occur in automatic mechanical and electrical systems.
To compensate for harmonics of a periodic disturbance, time delays are essential.
Repetitive control (\cite{1988_Hara_RC}), iterative learning control (\cite{2009_WANG_RC,2006_Bristow_RC}), and periodic disturbance observers (\cite{2018_Muramatsu_APDOB}) have been proposed using time delays for an internal model of a periodic disturbance.
However, in designing the sensitivity function with time delays, a waterbed-effect-like tradeoff exists between the wideband harmonic suppression, amplification of aperiodic disturbances, and deviation of harmonic suppression frequencies (\cite{2008_PipeleersRC,2014_Chen_DOBbasedRC,2021_Nie_DOBbasedRC,2023_Tanaka_PDOB,2023_Yang_PDOB,2021_Lai_PDOB,2023_Li_PDOB}).
To overcome the tradeoff, \cite{Muramatsu2025TCST} proposed a quasiperiodic disturbance observer (QDOB) using time delays, which achieved wideband harmonic suppression, non-amplification of aperiodic disturbances, and non-deviation of harmonic suppression frequencies, simultaneously.
From the perspective of the Bode sensitivity integral, this paper examines the waterbed effect on the QDOB in both continuous-time and discrete-time representations.
The Bode sensitivity integral for disturbance observers based on rational transfer functions has been studied by \cite{Sariyildiz2013icm,Sariyildiz2013iecon}.
However, the results cannot be applied to the QDOB by \cite{Muramatsu2025TCST}, whose open-loop transfer function is
\begin{align}
	\label{eq:CT:Gamma}
	\Gamma(s)&=\frac{\omega_\mathrm{c} L}{2}\frac{1+\Phi(s)}{1-\Phi(s)}\frac{\omega_\mathrm{b}}{s+\omega_\mathrm{b}},
\end{align}
where $\omega_\mathrm{b}, \omega_\mathrm{c}, L \in \mathbb{R}_{>0}$.
The function $\Phi(s)$ is a linear-phase low-pass filter that can be expressed as
\begin{align}
	\label{eq:Phi:delay}
	&\Phi(s)=e^{-\kappa T s}\prod_{i=1}^{l}\sum_{n=-N}^{N} \frac{\alpha_i(n)}{\gamma_i} e^{(n-N)T\bar{U}_is},
\end{align}
where $\kappa\ceq \bar{L}-N\sum_{i=1}^l\bar{U}_i\geq 1$, $\bar{U}_i\ceq \mathrm{round}(U_i/T)$, and $\bar{L}\ceq \mathrm{round}(L/T)$.
The open-loop transfer function has an infinite number of zeros by $1+\Phi(s)$ and an infinite number of poles by $1-\Phi(s)$ on the imaginary axis.
Consequently, the open-loop transfer function $\Gamma(s)$ is not rational and is a meromorphic function with time delays.
Several studies have extended the class of open-loop transfer functions applicable to the Bode sensitivity integral.
\cite{Freudenberg1987} investigated the Bode sensitivity integral with an open-loop transfer function $\Gamma_0(s)e^{-\tau s}$ combining a rational function $\Gamma_0(s)$ and time delay $e^{-\tau s}$.
In the discrete-time domain, the integral with iterative learning control was examined by \cite{songchon2000iterative}.
Besides, \cite{CHANG20234331} considered fractional-order proportional-integral-derivative control, while this paper does not consider such irrational fractional-order open-loop transfer functions.
Furthermore, \cite{Gomez1998} obtained the Bode sensitivity integral for meromorphic functions with time delays; however, it cannot address an infinite number of zeros of the sensitivity function on the imaginary axis, such as the QDOB.
The main results of this paper are the theorems of Bode-like sensitivity integrals for the QDOB in both continuous-time and discrete-time representations.
Through the theorems and numerical integration results, this paper demonstrates how the QDOB avoids the waterbed effect.

\section{Preliminaries} \label{sec:2}
This section presents the QDOB proposed by \cite{Muramatsu2025TCST} as preliminaries.
Suppose the single-input-single-output plant
\begin{align}
	\label{eq:plant}
	\mathcal{L}[y(t)]=P(s)\mathcal{L}[u(t)+v(t)],
\end{align}
where
\begin{subequations}
	\label{eq:Pdelta}
\begin{align}
	\label{eq:}
	P(s)&\ceq (1+\Delta(s))P_\mathrm{n}(s)\\
	P_\mathrm{n}(s)&\ceq\frac{b_{n}s^{n}+b_{n-1}s^{n-1}+\cdots+b_{1}s+b_0}{s^{m}+a_{m-1}s^{m-1}+\cdots+a_{1}s+a_0}\\
	\Delta(s)&\ceq\frac{\beta_{g}s^{g}+\beta_{g-1}s^{g-1}+\cdots+\beta_{1}s+\beta_0}{s^{h}+\alpha_{h-1}s^{h-1}+\cdots+\alpha_{1}s+\alpha_0}e^{-\gamma s}
\end{align}
\end{subequations}
with $n, m, g, h\in \mathbb{Z}_{\geq0}$, $n<m$, and $\gamma\in \mathbb{R}_{\geq0}$.
The symbols $t\in \mathbb{R}$, $u(t)\in \mathbb{R}$, $v(t)\in \mathbb{R}$, $y(t)\in \mathbb{R}$, $P(s)\in \mathbb{C}$, $P_\mathrm{n}(s)\in \mathbb{C}$, $\Delta(s)\in \mathbb{C}$, and $\mathcal{L}$ denote the time, input, exogenous signal, output, plant, nominal plant, stable modeling error, and Laplace transform operator, respectively.
Assume that the strictly proper transfer function $P_\mathrm{n}(s)$ has no zeros or poles in the closed right-half plane.
For the plant, the disturbance $d(t)$ including the exogenous signal and effects of the modeling error is defined as
\begin{align}
	\label{eq:}
	\mathcal{L}[d(t)] \ceq \mathcal{L}[v(t)] + \Delta(s)\mathcal{L}[u(t)+v(t)],
\end{align}
and the plant is rewritten as
\begin{align}
	\label{eq:sys:yud}
	\mathcal{L}[y(t)]=P_\mathrm{n}(s)\mathcal{L}[u(t)+d(t)].
\end{align}
Using the lifting transform $\mathcal{G}$ and the discrete-time Fourier transform $\mathcal{F}$:
\begin{subequations}
	\label{eq:}
\begin{align}
	&\mathcal{G}[d]_{L,\tau}(c) \ceq d(cL+\tau),\ L\in \mathbb{R}_{>0},\ \tau\in [0,L),\ c\in \mathbb{Z}\\
	&\mathcal{F}[D](\Omega)\ceq \sum_{c=-\infty}^\infty D(c) e^{-j\Omega c},
\end{align}
\end{subequations}
where $\Omega\in[0,2\pi)$ [rad/sample] denotes the normalized angular frequency.
Then, the definition of a quasiperiodic disturbance given in \cite{Muramatsu2025TCST} can be stated as Definition~\ref{def:QPdis}.
\begin{defn}
	\label{def:QPdis}
	A disturbance $d\in\mathcal{D}_L$ is called quasiperiodic with respect to a period $L\in \mathbb{R}_{>0}$ and separation frequency $\rho\in[0,\pi/L)$ if $d\in\mathcal{P}_{L,\rho} \subset \mathcal{D}_L$.
	\begin{subequations}
	\begin{align}
		\mathcal{D}_L&\ceq \{d:\mathbb{R}\to\mathbb{R}\mid\mathcal{G}[d]_{L,\tau}\in \ell^1(\mathbb{Z}),\ \forall\tau\in[0,L)\}\\
		\label{def:QPdis:PLp}
		\mathcal{P}_{L,\rho} &\ceq \{
			d\in\mathcal{D}_L\mid
			\mathcal{F}[\mathcal{G}[d]_{L,\tau}](\Omega)=0,\notag\\
			&\hspace{6em}\forall \tau\in [0,L),\ \forall\Omega\in(\rho L,\pi]
		\}.
	\end{align}
	\end{subequations}
\end{defn}
This definition is based on the quasiperiodicity in \cite{2022_Muramatsu_PA}.
Let $\omega_0\ceq 2\pi/L$ and $n\omega_0$ be the fundamental and harmonic frequencies of the quasiperiodic disturbance, respectively.
The QDOB, which estimates and compensates for the quasiperiodic disturbance, is
\begin{subequations}
	\label{eq:QDOB}
\begin{align}
	&\mathcal{L}[\hat{d}(t)]= Q(s) \mathcal{L}[\xi(t)-u(t)]\\
	&\mathcal{L}[\xi(t)]= B(s)P_\mathrm{n}^{-1}(s)\mathcal{L}[y(t)]\\
	&u(t)=r(t)-\hat{d}(t),
\end{align}
\end{subequations}
where $\hat{d}(t)\in \mathbb{R}$ is the estimated quasiperiodic disturbance.
The QDOB is composed of a first-order low-pass filter
\begin{align}
	\label{eq:}
	B(s) \ceq \frac{\omega_\mathrm{b}}{s+\omega_\mathrm{b}}
\end{align}
and the Q-filter
\begin{subequations}
	\label{eq:}
\begin{align}
	Q(s)&\ceq\frac{\omega_\mathrm{c} L(1+\Phi(s))}{(\omega_\mathrm{c} L+2)+(\omega_\mathrm{c} L-2)\Phi(s)}\\
	\label{def:QDOB:wc}
	\omega_\mathrm{c}&\ceq\frac{2}{L}\tan{\left(\frac{L}{2}\rho\right)},
\end{align}
\end{subequations}
using the linear-phase low-pass filter $\Phi(s)$:
\begin{subequations}
	\label{eq:QDOB:CT:Phi}
\begin{align}
	\label{def:CT:Phi}
	&\Phi(s)\ceq e^{-\kappa Ts}\prod_{i=1}^{l}\varphi_i(s)\\
	\label{def:CT:varphi}
	&\varphi_i(s) \ceq\frac{1}{\gamma_i}\sum_{n=-N}^{N}\alpha_i(n)e^{(n-N)T\bar{U}_is}\\
	&\alpha_i(n) \ceq w(n,N)h(n,\omega_i,U_i)\\
	\label{def:CT:Phi:w}
	&w(n,N)\ceq\left\{
	\begin{array}{cl}
		\multicolumn{2}{l}{0.42+0.5\cos({n\pi}{/N})}\\
		+0.08\cos({2n\pi}/{N})&\mathrm{if}\ |n|\leq N\\
		0&\mathrm{if}\ |n|>N,
	\end{array}
	\right.\\
	&h(n,\omega_i,U_i)\ceq\left\{
	\begin{array}{cl}
		{U_i\omega_i}/{\pi}&\mathrm{if}\ n=0\\
		{\sin(nU_i\omega_i)}/{(n\pi)}&\mathrm{if}\ n\neq0\\
	\end{array}
	\right.\\
	\label{def:CT:gamma}
	&\gamma_i\ceq\max_{\Omega\in [0,\pi]}\bigg|\alpha_i(0)+2\sum_{n=1}^N \alpha_i(n)\cos(n\Omega) \bigg|\\
	\label{def:CT:Phi:Ui}
	&U_{i}\ceq\left\{
	\begin{array}{cl}
		T&\mathrm{if}\ i=1\\
		{\pi}/{\omega_{i-1}}&\mathrm{otherwise}
	\end{array}
	\right.\\
	\label{def:CT:Phi:omegai}
	&\omega_i\ceq\left\{
	\begin{array}{cl}
		\omega_\mathrm{a}&\mathrm{if}\ i=l\\
		{2c\pi}/{U_i}&\mathrm{otherwise},
	\end{array}
	\right.\
	c=\frac{1}{2}\left(\frac{T\omega_\mathrm{a}}{\pi}\right)^{1/l}\\
	&N\ceq\min\{\max(\mathcal{N}),\ N_\mathrm{max}\}\\
	\label{def:CT:Phi:calN}
	&\mathcal{N} \ceq \{n\in \mathbb{Z}_{>0} \mid n \leq (L-T)/\textstyle\sum_{i=1}^lU_i\}.
\end{align}
\end{subequations}
The parameters $T\in \mathbb{R}_{>0}$, $\omega_\mathrm{a}\in \mathbb{R}_{>0}$, $l\in \mathbb{Z}_{>0}$, and $N_\mathrm{max}\in \mathbb{Z}_{>0}$ denote the sampling time, cutoff frequency, number of stages, and maximum order, respectively.
This linear-phase low-pass filter can be equivalently expressed as \eqref{eq:Phi:delay}.
The Q-filter is based on the periodic/aperiodic separation filter (\cite{2022_Muramatsu_PA,2019_Muramatsu_PASF}) whose time delays are multiplied by zero-phase low-pass filters.
Consequently, the open-loop transfer function $\Gamma(s)$ of the QDOB is
\begin{align}
	\Gamma(s)&=\frac{\omega_\mathrm{c} L}{2}\frac{1+\Phi(s)}{1-\Phi(s)}\frac{\omega_\mathrm{b}}{s+\omega_\mathrm{b}},\tag{\getrefnumber{eq:CT:Gamma}}
\end{align}
and its sensitivity and complementary sensitivity functions are
\begin{align}
	\label{eq:CT:SandT}
	S(s)&\coloneqq \frac{1}{1+\Gamma(s)},\
	T(s)\coloneqq \frac{\Gamma(s)}{1+\Gamma(s)},
\end{align}
respectively.
The QDOB is discretized by two methods.
The Q-filter $Q(s)$ is discretized by $e^{-Ts}\to z^{-1}$, and the transfer function $B(s)P_\mathrm{n}^{-1}(s)$ is discretized by the backward Euler method $s\to (1-z^{-1})/T$.
Then, the open-loop transfer function in the discrete-time representation is
\begin{align}
	\label{eq:DT:Gamma}
	\tilde{\Gamma}(z)&=\frac{\omega_\mathrm{c} L}{2}\frac{1+\tilde{\Phi}(z)}{1-\tilde{\Phi}(z)}\frac{\omega_\mathrm{b}Tz}{(1+\omega_\mathrm{b}T)z-1},
\end{align}
where
\begin{subequations}
	\label{eq:DT:Phi}
\begin{align}
	&\tilde{\Phi}(z)\ceq z^{-\kappa}\prod_{i=1}^{l}\tilde{\varphi}_i(z)\\
	&\tilde{\varphi}_i(z) \ceq\frac{\sum_{n=-N}^{N}w(n,N)h(n,\omega_i,U_i)z^{(n-N)\bar{U}_i}}{\sum_{n=-N}^{N}w(n,N)h(n,\omega_i,U_i)}.
\end{align}
\end{subequations}
Accordingly, the sensitivity and complementary sensitivity functions in the discrete-time representation are
\begin{align}
	\label{eq:DT:SandT}
	\tilde{S}(z)&\coloneqq \frac{1}{1+\tilde{\Gamma}(z)},\
	\tilde{T}(z)\coloneqq \frac{\tilde{\Gamma}(z)}{1+\tilde{\Gamma}(z)},
\end{align}
respectively.
%
%
%
\begin{figure}[t!]
		\begin{center}
			\includegraphics[width=0.9\hsize]{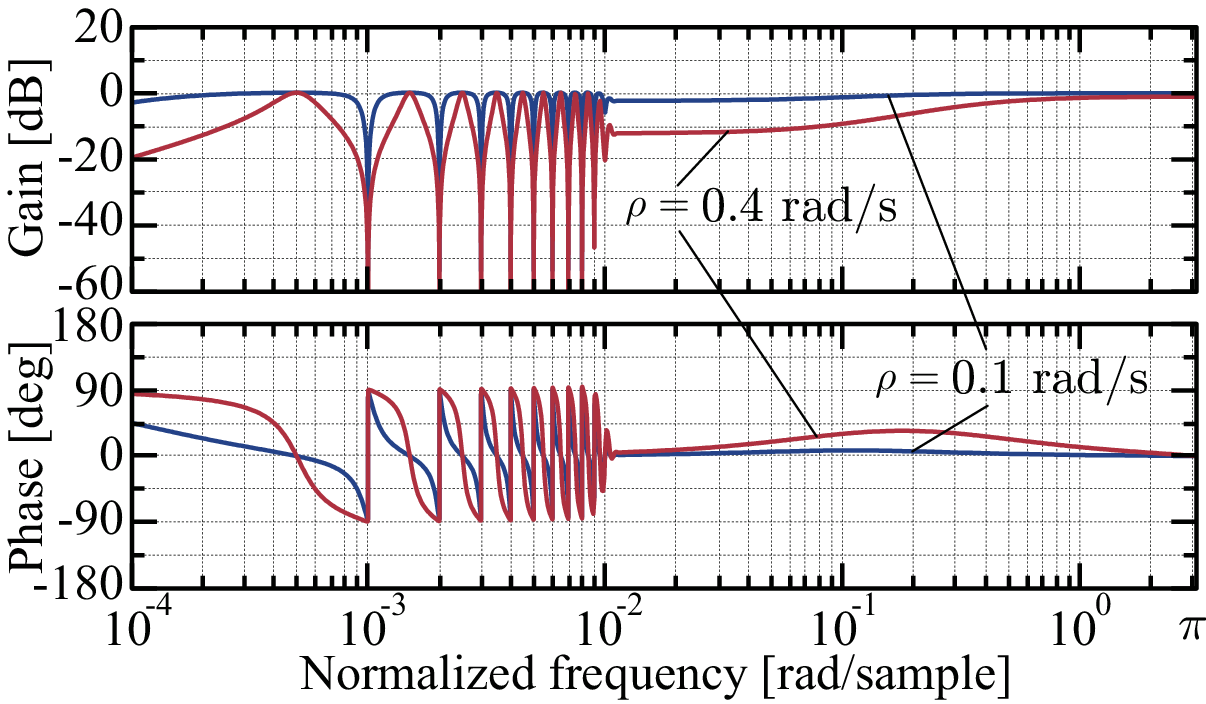}\\
			(a) Sensitivity function $\tilde{S}(e^{j\Omega})$ in \eqref{eq:DT:SandT}.\\
			\includegraphics[width=0.9\hsize]{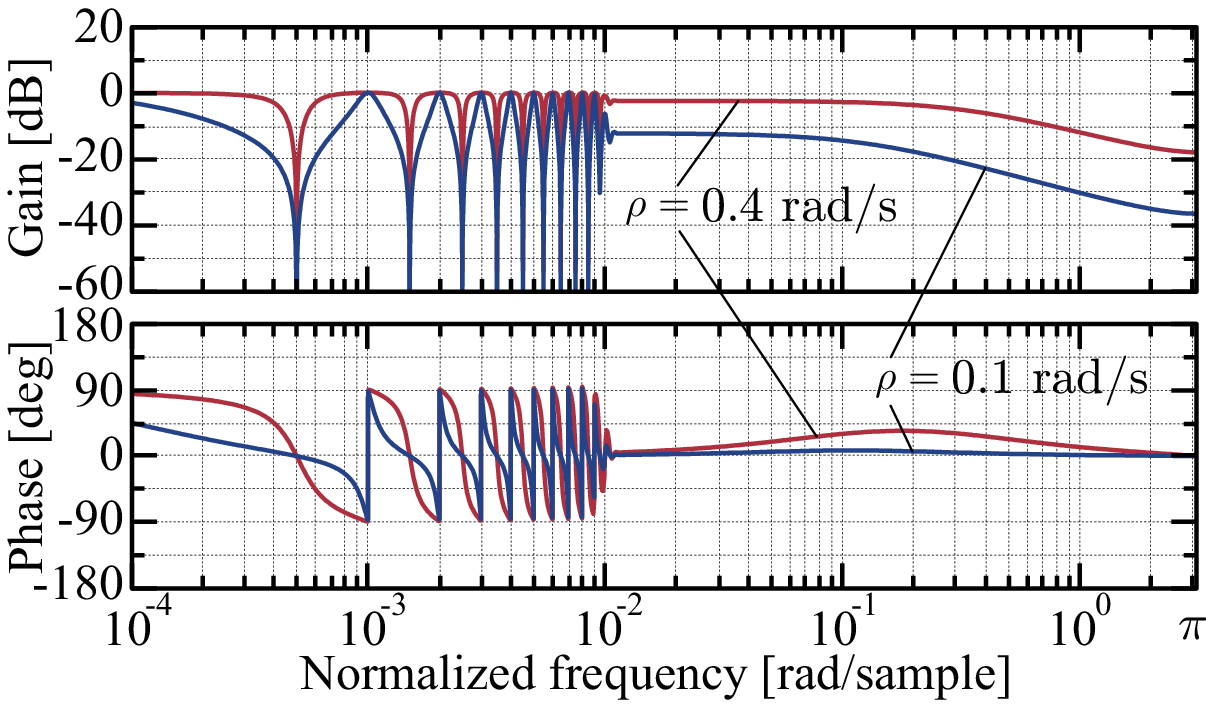}\\
			(b) Complementary sensitivity function $\tilde{T}(e^{j\Omega})$ in \eqref{eq:DT:SandT}.\\
			\includegraphics[width=0.9\hsize]{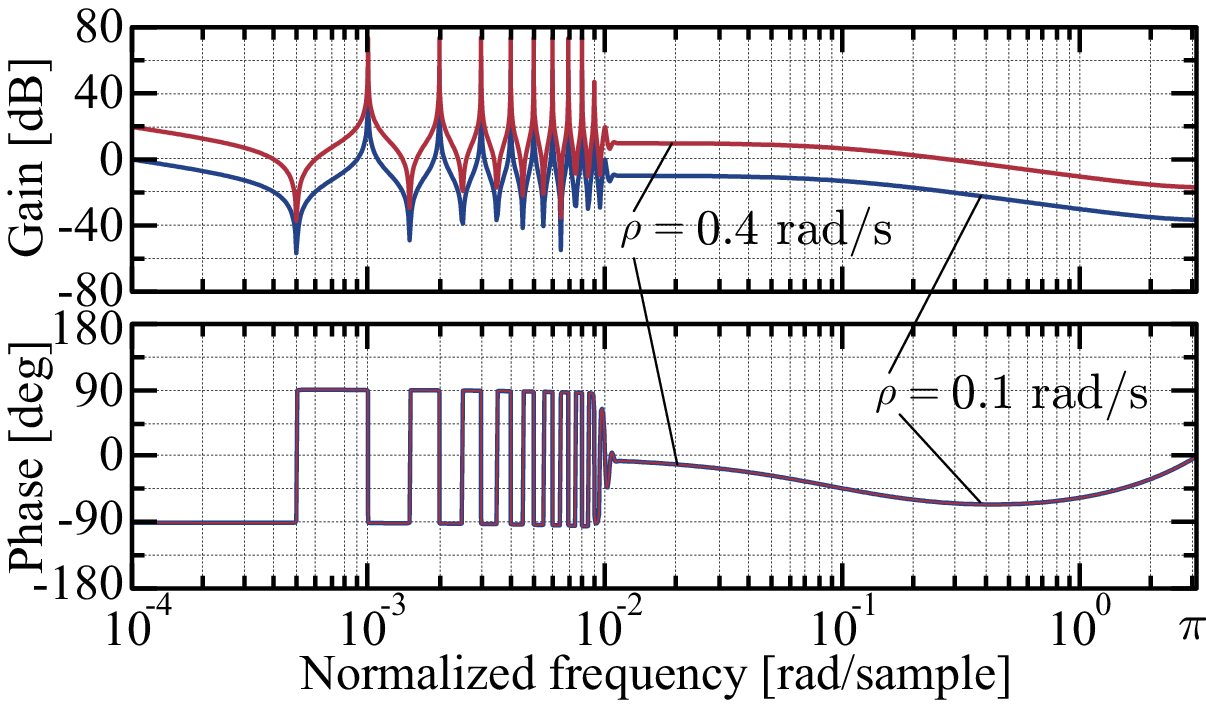}\\
			(c) Open-loop transfer function $\tilde{\Gamma}(e^{j\Omega})$ in \eqref{eq:DT:Gamma}.
		\end{center}
		\caption{Bode plots of the QDOB in the discrete-time representation with $l=3$, $N_\mathrm{max}=256$, $\omega_0=1$ rad/s, $\omega_\mathrm{a}=10$ rad/s, $\omega_\mathrm{b}=100$ rad/s, and $T=1$ ms.}\label{fig:BodePlots}
\end{figure}
Fig.~\ref{fig:BodePlots} shows the Bode plots of the sensitivity, complementary sensitivity, and open-loop transfer functions in the discrete-time representation.
As shown in Fig.~\ref{fig:BodePlots}(a), the QDOB can expand the harmonic suppression bandwidth by increasing the separation frequency $\rho$ without the waterbed effect.
See \cite{Muramatsu2025TCST} for the detailed derivation, design, analyses, and discretization of the QDOB, including the Bode plots of the functions in the continuous-time representation.
Remark~\ref{rem:change} describes the modifications to \cite{Muramatsu2025TCST}, and Remark~\ref{rem:design} introduces how to design the QDOB on the basis of \cite{Muramatsu2025TCST}.
\begin{rem}
	\label{rem:change}
	The normalization $1/\gamma_i$ of \eqref{def:CT:varphi} for the linear-phase low-pass filter $\Phi(s)$ has been modified so that the gain of the filter satisfies $|\Phi(j\omega)|\leq1$.
	In \cite{Muramatsu2025TCST}, it was normalized by the DC gain $\gamma_i=\alpha_i(0)+2\sum_{n=1}^N \alpha_i(n)$ by neglecting passband ripple of the filter.
	Additionally, rounded $\bar{U}_iT$ and $\bar{L}T$ are used for the continuous-time representation instead of $U_i$ and $L$ in the time delays.
	Lastly, minor errors in the set $(\rho L,\pi]$ for $\Omega$ of \eqref{def:QPdis:PLp} in Definition~\ref{def:QPdis} and of the upper limit of the summation of \eqref{def:CT:Phi:calN} have been corrected.
\end{rem}
\begin{rem}
	\label{rem:design}
	The crucial hyperparameters of the QDOB for the sensitivity design are $\omega_\mathrm{a}$, $\omega_\mathrm{b}$, and $\rho$.
	The sensitivity function suppresses harmonics at frequencies below $\omega_\mathrm{a}$, and its gain can be less than 1 at frequencies below $\omega_\mathrm{b}$.
	Additionally, the gain of the complementary sensitivity function decreases for robust stability at frequencies above $\omega_\mathrm{b}$.
	The angular frequencies $\omega_\mathrm{a}$ and $\omega_\mathrm{b}$ are adjusted to design the sensitivity and complementary sensitivity functions within the condition $\omega_\mathrm{a}\ll\omega_\mathrm{b}$, which maintains the phase of the open-loop transfer function within $-90$ deg. to $90$ deg. at frequencies below the Nyquist frequency.
	The separation frequency $\rho\in[0,\pi/L)$ changes the harmonic suppression bandwidth such that $20\log|S(j\omega)|<-3$ dB $\forall\omega\in [n\omega_0-\rho,n\omega_0+\rho]\cap[0,\omega_\mathrm{a}]$ around the harmonic frequencies $n\omega_0$.
\end{rem}
Based on the preliminaries, this paper discusses the waterbed effect on the sensitivity functions $S(s)$ and $\tilde{S}(z)$ in the continuous-time and discrete-time representations.

\section{Waterbed Effect on QDOB} \label{sec:3}
\subsection{Main Results}
\begin{figure}[t]
	\begin{center}
		\includegraphics[width=\hsize]{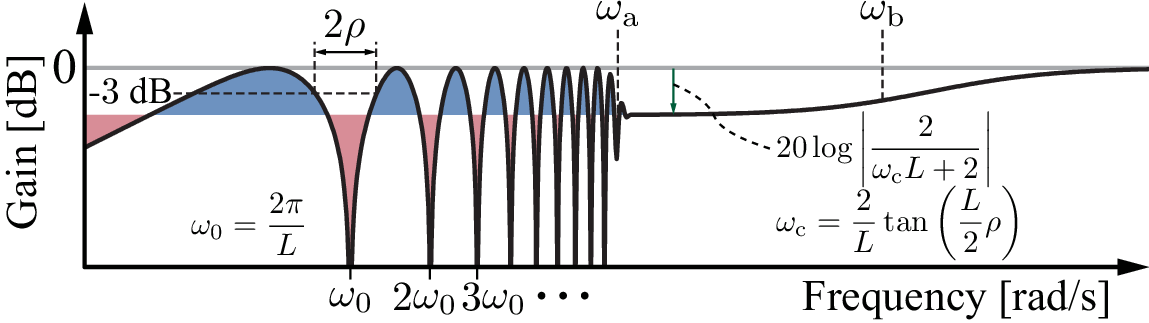}
	\end{center}
	\caption{Illustrative diagram for the gains of the sensitivity functions: $S(s)$ in \eqref{eq:CT:SandT} and $\tilde{S}(z)$ in \eqref{eq:DT:SandT} of the QDOB.}\label{fig:S-concept}
\end{figure}

The main results of this paper are two theorems that apply a Bode-like sensitivity integral to the QDOB in the continuous-time or discrete-time representation.
Two lemmas are provided for the proofs of the theorems.
\begin{lem}
	\label{lem:Phi}
	Suppose \eqref{eq:QDOB:CT:Phi} and $g\in \mathbb{R}_{\geq0}$; then,
	\begin{subequations}
		\label{eq:}
	\begin{align}
		&\mathcal{S}_g\ceq \{s\in \mathbb{C}_+ \mid \rmRe[s]>g\}\\
		&|\Phi(s)|<\max_{\rmRe[s]=g}|\Phi(s)|,\ \forall s\in \mathcal{S}_g.
	\end{align}
	\end{subequations}
\end{lem}
\begin{pf}
	The proof is in Appendix~\ref{sec:app:PHI}.
\end{pf}
\begin{lem}
	\label{lem:OLTF}
	The open-loop transfer function \eqref{eq:CT:Gamma} has no poles in the open right-half plane.
\end{lem}
\begin{pf}
	The proof is in Appendix~\ref{sec:app:OLTF}.
\end{pf}
\begin{thm}
	\label{thm:Bode:Integral:QDOB:CT}
	Suppose the open-loop transfer function \eqref{eq:CT:Gamma}.
	Then, the sensitivity function \eqref{eq:CT:SandT} in the continuous-time representation satisfies
	\begin{align}
		\label{eq:Bode:Integral:QDOB:CT}
		\lim_{\varepsilon\to0^+}\int_{0}^{\infty}\ln|S(\varepsilon+j\omega)|d\omega
		=-\frac{\pi\omega_\mathrm{b}\omega_\mathrm{c} L}{4}.
	\end{align}
\end{thm}
\begin{pf}
	The proof is in Appendix~\ref{sec:app:Bode:CT}.
\end{pf}
\begin{thm}
	\label{thm:Bode:Integral:QDOB:DT}
	Suppose the open-loop transfer function \eqref{eq:DT:Gamma}.
	Then, the sensitivity function \eqref{eq:DT:SandT} in the discrete-time representation satisfies
	\begin{align}
		\label{eq:Bode:Integral:QDOB:DT}
		&\lim_{\varepsilon\to1^+}\int_{0}^{2\pi}\ln|\tilde{S}(\varepsilon e^{j\Omega})|d\Omega\notag\\
		&=2\pi\ln(2+2\omega_\mathrm{b}T)
		-2\pi\ln(2+2\omega_\mathrm{b}T+\omega_\mathrm{b}\omega_\mathrm{c} LT).
	\end{align}
\end{thm}
\begin{pf}
	The proof is in Appendix~\ref{sec:app:Bode:DT}.
\end{pf}
From the theorems, the Bode-like sensitivity integrals \eqref{eq:Bode:Integral:QDOB:CT} and \eqref{eq:Bode:Integral:QDOB:DT} decrease in both continuous-time and discrete-time representations as $\omega_\mathrm{b}$ and/or $\rho$ increases.
This is consistent with Remark~\ref{rem:design}, the gain of the sensitivity function decreases as the parameters increase.
Note that $\omega_\mathrm{c}$ is determined by $\rho$ according to \eqref{def:QDOB:wc}, and they are in a monotonic relationship.
These suggest that the QDOB can extend the harmonic suppression bandwidth without a sensitivity tradeoff with the decrease in the integral.
In contrast, $\omega_\mathrm{a}$ does not affect the integrals.
This implies that the band stops around the harmonic frequencies below $\omega_\mathrm{a}$ are restricted by the sensitivity tradeoff.
Fig.~\ref{fig:S-concept} illustrates the relationship among the gain of the sensitivity function, $\omega_\mathrm{a}$, $\omega_\mathrm{b}$, $\omega_\mathrm{c}$, and $\rho$.
It shows that $\omega_\mathrm{b}$ extends the suppression bandwidth horizontally, and $\omega_\mathrm{c}L$ deepens the attenuation depth vertically.
Meanwhile, there exists the waterbed effect with $\omega_\mathrm{a}$ as the red region of the reduced gain and the blue region of the increased gain, while the gain does not exceed 0 dB.
%

%
Theorems~\ref{thm:Bode:Integral:QDOB:CT} and \ref{thm:Bode:Integral:QDOB:DT} employ the Bode-like sensitivity integrals based on the limits avoiding the singularities on the imaginary axis and unit disk because the sensitivity function has an infinite and finite number of roots, respectively.
Although the QDOB does not satisfy the assumptions of the existing Bode sensitivity integrals, the Bode-like sensitivity integrals \eqref{eq:Bode:Integral:QDOB:CT} and \eqref{eq:Bode:Integral:QDOB:DT} turn out to be consistent with the right-hand sides of the Bode sensitivity integrals \eqref{eq:Bode:integral:conv:CT} and \eqref{eq:Bode:integral:conv:DT}, respectively.
%

%
\begin{table}[t]
	\caption{Seven parameter settings and theoretical values of the Bode-like sensitivity integrals $\delta_\mathrm{ct}$ and $\delta_\mathrm{dt}$ in \eqref{eq:delta} with $T=1$ ms.}\label{tab:para}
	\begin{center}
		\begin{tabular}{ccccccc}
			\hline
			&$\omega_0$&$\omega_\mathrm{b}$&$\rho$&$\delta_\mathrm{ct}$&$\delta_\mathrm{dt}$\\
			\hline \hline
			$\mathrm{P}_1$ &1 rad/s&100 rad/s&0.1 rad/s& -51.0& -0.18\\
			$\mathrm{P}_2$&1 rad/s&100 rad/s&0.25 rad/s&-157.1& -0.55 \\
			$\mathrm{P}_3$&10 rad/s&400 rad/s&1.476 rad/s&-314.2& -0.84 \\
			$\mathrm{P}_4$&1 rad/s&200 rad/s&0.25 rad/s&-314.2& -0.97 \\
			$\mathrm{P}_5$&1 rad/s&0.1 rad/s&0.25 rad/s&-0.16&-0.63$\times10^{-3}$\\
			$\mathrm{P}_6$&1 rad/s&0.5 rad/s&0.25 rad/s&-0.79&-3.14$\times10^{-3}$\\
			$\mathrm{P}_7$&1 rad/s&1 rad/s&0.25 rad/s&-1.57&-6.27$\times10^{-3}$\\
			\hline
		\end{tabular}
	\end{center}
\end{table}
%
%

%
%
\subsection{Numerical Results}
Numerical examples are provided to verify the convergence of the integrals: $\int_{0}^{w}\ln|S(j\omega)|d\omega$ and $\int_{0}^{w}\ln|\tilde{S}(e^{j\Omega})|d\Omega$ with respect to the upper limit of integration $w$ toward \eqref{eq:Bode:Integral:QDOB:CT} and \eqref{eq:Bode:Integral:QDOB:DT}.
For the validation, parameter settings of Table~\ref{tab:para} for the hyperparameters: $\omega_0$, $\omega_\mathrm{b}$, and $\rho$ are considered.
The theoretical values from \eqref{eq:Bode:Integral:QDOB:CT} and \eqref{eq:Bode:Integral:QDOB:DT} are expressed as
\begin{subequations}
	\label{eq:delta}
\begin{align}
	\delta_\mathrm{ct} &\ceq - \frac{\pi\omega_\mathrm{b}\omega_\mathrm{c} L}{4}
	= - \frac{\pi\omega_\mathrm{b}}{2}\tan{\left(\frac{\pi\rho}{\omega_0}\right)}\\
	\delta_\mathrm{dt} &\ceq 2\pi\ln(2+2\omega_\mathrm{b}T) -2\pi\ln(2+2\omega_\mathrm{b}T+\omega_\mathrm{b}\omega_\mathrm{c} LT)\notag\\
	&=2\pi\ln(1+\omega_\mathrm{b}T)\notag\\
	&\hspace{2em} -2\pi\ln\left(1+\omega_\mathrm{b}T+ \omega_\mathrm{b}T \tan{(\pi\rho/\omega_0)}\right).
\end{align}
\end{subequations}
%

%
\begin{figure}[t!]
		\begin{center}
			\includegraphics[width=0.9\hsize]{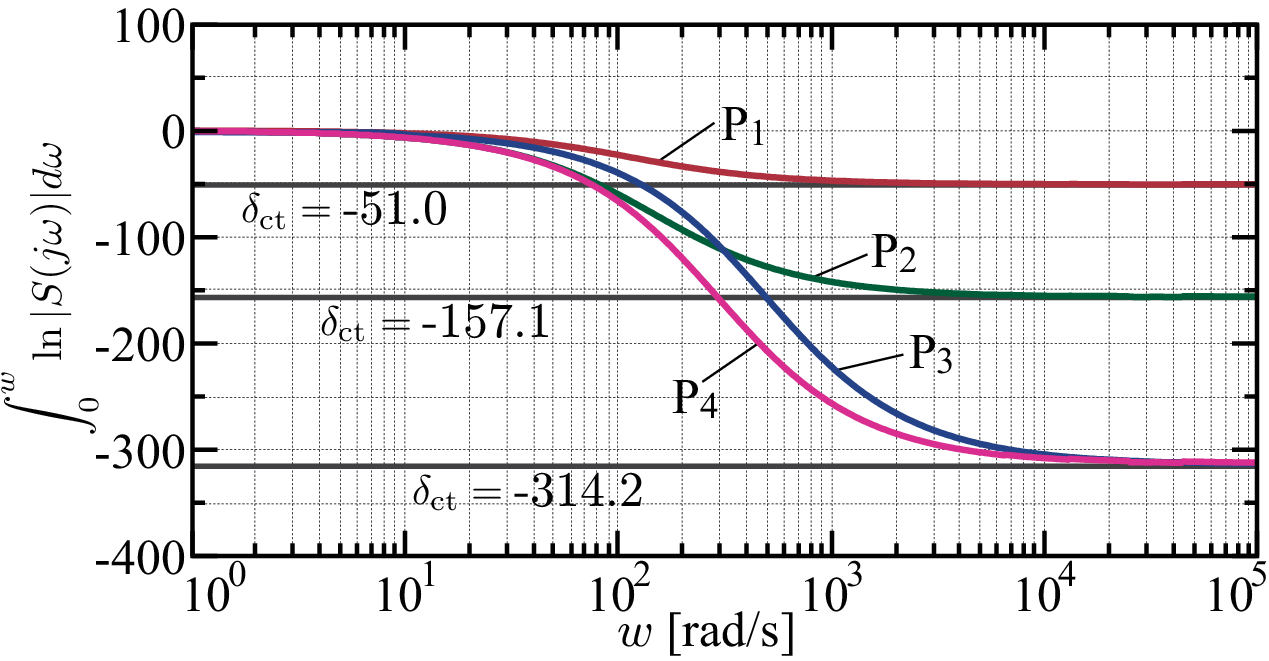}\\
			(a) Cases of $\omega_\mathrm{a}\ll\omega_\mathrm{b}$.\\
			\includegraphics[width=0.9\hsize]{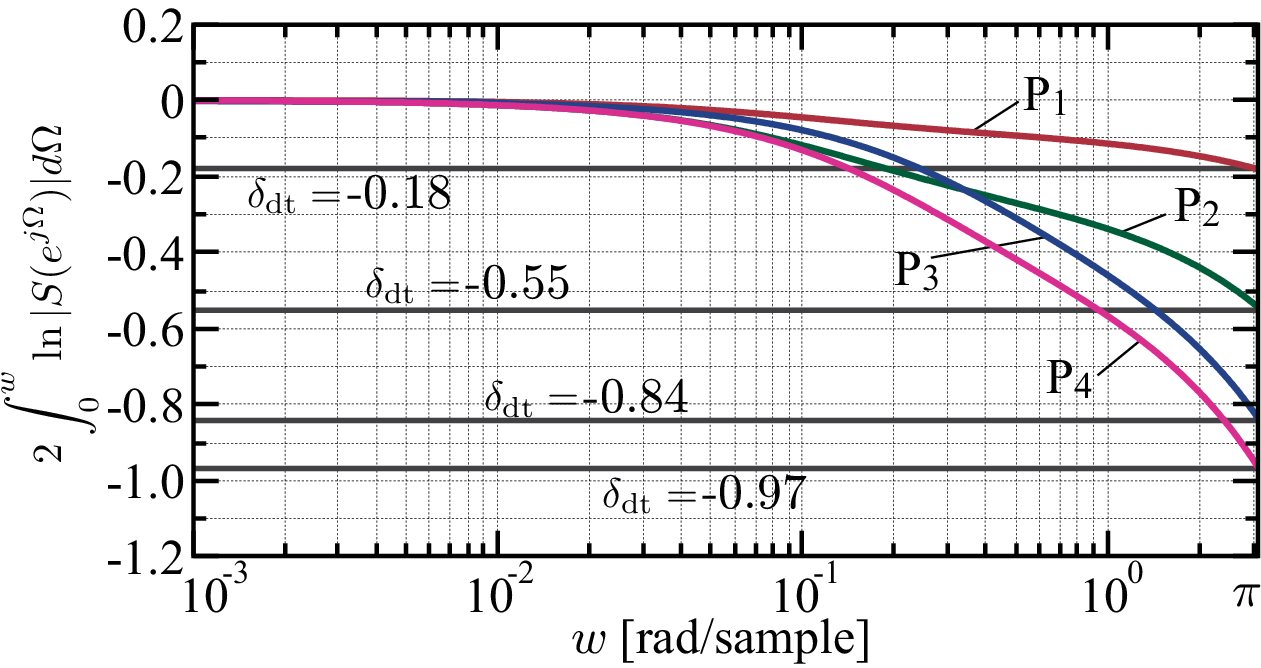}\\
			(b) Cases of $\omega_\mathrm{a}\ll\omega_\mathrm{b}$.\\
			\includegraphics[width=0.9\hsize]{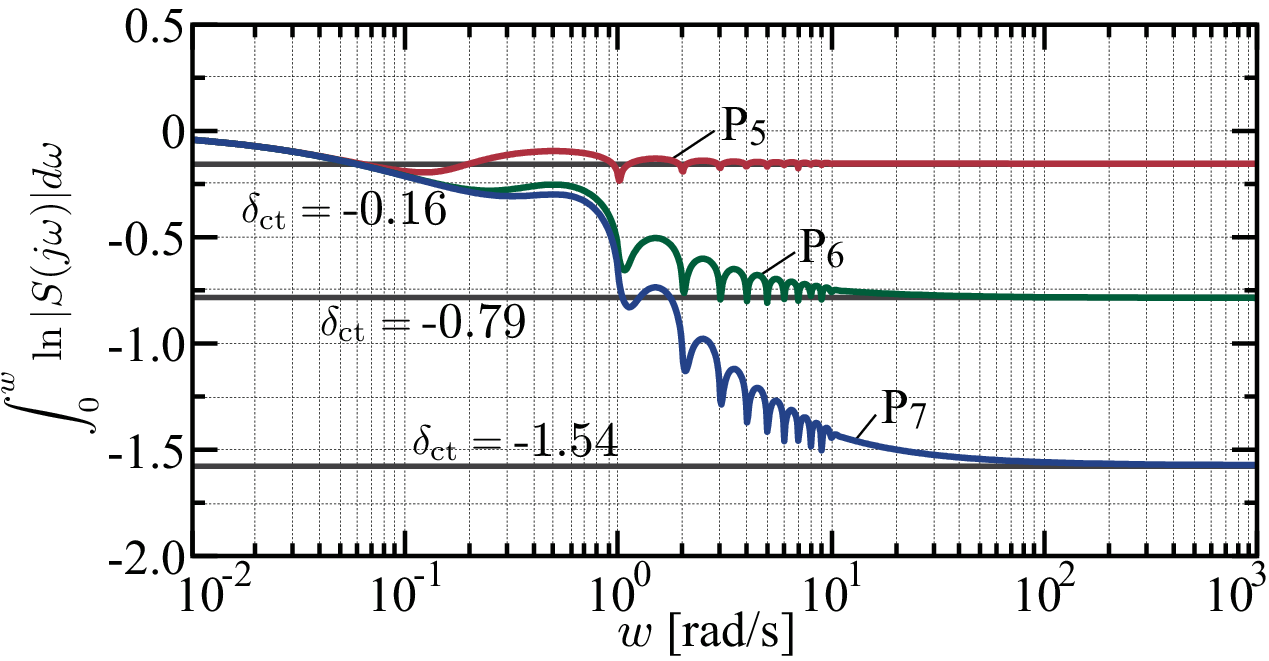}\\
			(c) Cases of $\omega_\mathrm{a}>\omega_\mathrm{b}$.\\
			\includegraphics[width=0.9\hsize]{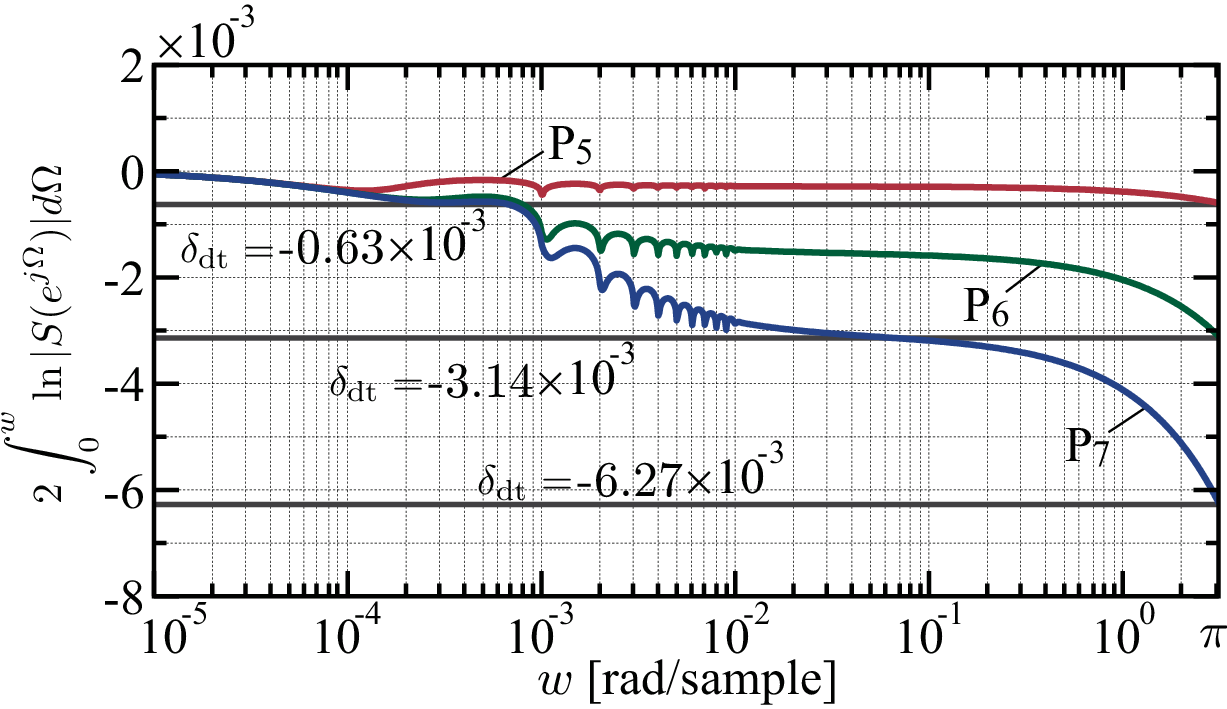}\\
			(d) Cases of $\omega_\mathrm{a}>\omega_\mathrm{b}$.
		\end{center}
		\caption{Convergence of the numerical Bode-like sensitivity integrals toward the theoretical values $\delta_\mathrm{ct}$ and $\delta_\mathrm{dt}$ in \eqref{eq:delta} with $l=3$, $N_\mathrm{max}=256$, $\omega_\mathrm{a}=10$ rad/s, $T=1$ ms, and Table~\ref{tab:para}.}\label{fig:NR}
\end{figure}
The numerical results are shown in Fig.~\ref{fig:NR}.
In the continuous-time cases of Fig.~\ref{fig:NR}(a) and (c), the integral values asymptotically converge toward $\delta_\mathrm{ct}$ as $w$ increases.
In the discrete-time cases of Fig.~\ref{fig:NR}(b) and (d), the integral values reach $\delta_\mathrm{dt}$ at $w=\pi$.
Note that
\begin{align}
	&\int_{0}^{2\pi}\ln|\tilde{S}(e^{j\Omega})|d\Omega=2\int_{0}^{\pi}\ln|\tilde{S}(e^{j\Omega})|d\Omega
\end{align}
based on $|\tilde{S}(e^{j\Omega})|=|\tilde{S}(e^{-j\Omega})|$ and $|\tilde{S}(e^{j\Omega})|=|\tilde{S}(e^{j(\Omega+2\pi)})|$.
Although Theorems~\ref{thm:Bode:Integral:QDOB:CT} and \ref{thm:Bode:Integral:QDOB:DT} hold regardless of the relationship between $\omega_\mathrm{a}$ and $\omega_\mathrm{b}$, the convergences of the integral values are different between $\omega_\mathrm{a}\ll\omega_\mathrm{b}$ [Fig.~\ref{fig:NR}(a) and (b)] and $\omega_\mathrm{a}>\omega_\mathrm{b}$ [Fig.~\ref{fig:NR}(c) and (d)].
If $\omega_\mathrm{a}>\omega_\mathrm{b}$, the gains $|S(j\omega)|$ and $|\tilde{S}(e^{j\Omega})|$ repeatedly increase and decrease while converging.
These increases correspond to the amplification of aperiodic disturbances and can cause a shift in harmonic suppression frequencies, which usually happen with conventional repetitive control.
Thus, the avoidance of the sensitivity tradeoff in Theorems~\ref{thm:Bode:Integral:QDOB:CT} and \ref{thm:Bode:Integral:QDOB:DT} does not prevent local amplification of the gain.
To avoid the local amplification, the phase of the open-loop transfer function within $\pm90$ deg. is essential.
To discuss this, this paper considers the following proposition.
\begin{prop}
	\label{prop:phase_gain}
	Suppose an open-loop transfer function $\Gamma(j\omega)$ and sensitivity function $S(j\omega)=(1+\Gamma(j\omega))^{-1}$. Then, $|\angle\Gamma(j\omega)|\leq \pi/2 \Rightarrow |S(j\omega)|\leq 1$.
\end{prop}
\begin{pf}
	$
	|\angle\Gamma(j\omega)|\leq \pi/2
	\Rightarrow \mathrm{Re}[\Gamma(j\omega)]\geq0
	\Rightarrow (1+\mathrm{Re}[\Gamma(j\omega)])^2+\mathrm{Im}[\Gamma(j\omega)]^2\geq1
	\Rightarrow |S(j\omega)|=((1+\mathrm{Re}[\Gamma(j\omega)])^2+\mathrm{Im}[\Gamma(j\omega)]^2)^{-1/2}\leq 1
	$.
\end{pf}
On the basis of Remark~\ref{rem:design} and Proposition~\ref{prop:phase_gain}, the parameter design $\omega_\mathrm{a}\ll\omega_\mathrm{b}$, which makes the phase of the open-loop transfer function within $\pm90$ deg. at frequencies below the Nyquist frequency, prevents local amplification of the gain at the frequencies.
Furthermore, the separation frequency $\rho$ does not affect the phase of the open-loop transfer function $\Gamma(s)$, and the phase exists within $\pm 90$ deg. regardless of $\rho$, as shown in Fig.~\ref{fig:BodePlots}(c).
This indicates that the gain of the sensitivity function always satisfies $|S(j\omega)|\leq 1$ regardless of $\rho$ if $\omega_\mathrm{a}\ll\omega_\mathrm{b}$.
These results explain how the increase in the separation frequency $\rho$ extends the harmonic suppression bandwidth while decreasing the Bode-like sensitivity integral, without causing amplification at other frequencies below the Nyquist frequency.
For the continuous-time representation, at frequencies beyond the Nyquist frequency, the low-pass filter $B(j\omega)$ and sensitivity function $S(j\omega)$ asymptotically approach zero and one, respectively, as the frequency increases.

\section{Conclusion}
This paper examined the waterbed effect in the QDOB proposed by \cite{Muramatsu2025TCST}, which estimates and compensates for the quasiperiodic disturbance of Definition~\ref{def:QPdis}.
Compared to conventional periodic-disturbance-compensation control, the advantage of the QDOB is the simultaneous realization of the wideband harmonic suppression, non-amplification of aperiodic disturbances, and proper harmonic suppression frequencies.
In this paper, Theorems~\ref{thm:Bode:Integral:QDOB:CT} and \ref{thm:Bode:Integral:QDOB:DT} indicated the avoided sensitivity tradeoff, where the extension of harmonic suppression bandwidth for the sensitivity function also reduces the Bode-like sensitivity integral.
Moreover, the numerical results and Proposition~\ref{prop:phase_gain} showed the waterbed effect on the QDOB, where the phase of the open-loop transfer function within $\pm 90$ deg. results in the non-amplification of aperiodic disturbances and proper harmonic suppression frequencies.
This clarified why the parameter condition $\omega_\mathrm{a}\ll\omega_\mathrm{b}$ in Remark~\ref{rem:design} is necessary for the avoidance of the waterbed effect.
These results support the advantage of the QDOB and clarify its practical potential for improving automatic mechanical and electrical systems with quasiperiodic, periodic, or harmonic disturbances, such as friction or gravitational force with repetitive motion, wind disturbances, torque ripple, and current ripple.
However, there remain two problems: how to find $\gamma_i$ defined in \eqref{def:CT:gamma} for implementation, and the waterbed effect associated with discretization methods.

\appendix
\section{Proof of Lemma~\getrefnumber{lem:Phi}}\label{sec:app:PHI}
The linear-phase low-pass filter $\Phi(s)$ in \eqref{eq:QDOB:CT:Phi} is analytic and can be expressed as \eqref{eq:Phi:delay}.
Let
\begin{subequations}
	\label{eq:}
\begin{align}
	&\mathcal{E}_{g}\ceq\{\xi\in \mathbb{C}\mid |\xi|< e^{-Tg}\},\
	\xi\ceq e^{-Ts}\\
	&H(\xi)\ceq \prod_{i=1}^{l}\sum_{n=-N}^{N} \frac{\alpha_i(n)}{\gamma_i} \xi^{(N-n)\bar{U}_i}.
\end{align}
\end{subequations}
Based on \eqref{eq:Phi:delay},
\begin{subequations}
	\label{eq:app1:xi}
\begin{align}
	&\xi \in \mathcal{E}_{g} \Leftrightarrow s\in \mathcal{S}_g,\
	|\xi|= e^{-Tg}\Leftrightarrow \rmRe[s]= g\\
	&\Phi(s)=e^{-\kappa Ts}H(e^{-Ts}).
\end{align}
\end{subequations}
Since $H(\xi)$ is nonconstant and analytic in the open disk $\mathcal{E}_{g}$ and continuous on the closure of $\mathcal{E}_{g}$, the maximum modulus principle gives
\begin{align}
	\label{eq:}
	|H(\xi)|<\max_{|\xi|=e^{-Tg}}|H(\xi)|,\ \forall\xi \in \mathcal{E}_g.
\end{align}
By the change of variable $\xi=e^{-Ts}$,
\begin{subequations}
	\label{eq:}
\begin{align}
	&|H(e^{-Ts})|=|e^{\kappa Ts}| |\Phi(s)|=e^{\kappa T\rmRe[s]}|\Phi(s)|\\
	&\max_{|\xi|=e^{-Tg}}|H(\xi)|=\max_{\rmRe[s]=g}|H(e^{-Ts})|\notag\\
	&\hspace{2em}=\max_{\rmRe[s]=g}|e^{\kappa Ts}\Phi(s)|=e^{\kappa Tg}\max_{\rmRe[s]=g}|\Phi(s)|.
\end{align}
\end{subequations}
Hence,
\begin{align}
	&e^{\kappa T\rmRe[s]} |\Phi(s)|<e^{\kappa Tg}\max_{\rmRe[s]=g}|\Phi(s)|,\ \forall s\in \mathcal{S}_g.
\end{align}
Because $e^{\kappa T\rmRe[s]}>e^{\kappa Tg}\geq1$, $\forall s\in \mathcal{S}_g$, we have
\begin{align}
	\label{eq:}
	|\Phi(s)|<\max_{\rmRe[s]=g}|\Phi(s)|,\ \forall s\in \mathcal{S}_g.
\end{align}

\section{Proof of Lemma~\getrefnumber{lem:OLTF}}\label{sec:app:OLTF}
The poles of \eqref{eq:CT:Gamma} are roots of $s+\omega_\mathrm{b}$ and $1-\Phi(s)$.
The root of $s+\omega_\mathrm{b}$ is $s=-\omega_\mathrm{b}$ in the open left-half plane.
The linear-phase low-pass filter $\Phi(s)$ in \eqref{eq:QDOB:CT:Phi} consists of multiple linear-phase low-pass filters $\varphi_i(s)$ in \eqref{def:CT:varphi}.
Its gain is
\begin{align}
	\label{eq:}
	|\varphi_i(j\omega)|&=\frac{1}{\gamma_i}\left|
		\alpha_i(0) + \sum_{n=1}^{N}\alpha_i(n)(e^{jnT\bar{U}_i\omega}+e^{-jnT\bar{U}_i\omega})
	\right|\notag\\
	&=\frac{1}{\gamma_i}\left|
		\alpha_i(0) + 2\sum_{n=1}^{N}\alpha_i(n)\cos(nT\bar{U}_i\omega)
	\right|
\end{align}
using the property $\alpha_i(n)=\alpha_i(-n)$ of even functions.
Eqs. \eqref{def:CT:varphi} and \eqref{def:CT:gamma} yield $|\varphi_i(j\omega)|\leq1$, $\forall\omega\in \mathbb{R}$, and
\begin{align}
	\label{eq:app2:absPhiOmega}
	\left|\Phi(j\omega)\right|
	=\prod_{i=1}^{l}\left|\varphi_i(j\omega)\right|
	\leq1,\ \forall\omega\in \mathbb{R}.
\end{align}
Lemma~\ref{lem:Phi} gives
\begin{align}
	\label{eq:}
	|\Phi(s)|<\max_{\rmRe[s]=0}|\Phi(s)|,\ \forall s\in \mathcal{S}_0= \mathbb{C}_+.
\end{align}
These yield
\begin{align}
	\label{eq:app2:absPhi:RHP}
	|\Phi(s)|<\max_{\omega\in \mathbb{R}}|\Phi(j\omega)|\leq 1,\ \forall s\in \mathbb{C}_+,
\end{align}
which implies $\Phi(s)\neq 1,\ \forall s\in \mathbb{C}_+$, and $1-\Phi(s)$ has no roots in $\mathbb{C}_+$.
Therefore, the open-loop transfer function \eqref{eq:CT:Gamma} has no poles in the open right-half plane.

\section{Proof of Theorem~\getrefnumber{thm:Bode:Integral:QDOB:CT}}\label{sec:app:Bode:CT}
Let a simply connected region $\mathcal{C}_{\varepsilon}\subset\mathbb{C}_+$ and simple closed contour $\mathcal{D}_{\varepsilon,\delta}\subset\mathcal{C}_{\varepsilon}$ be
\begin{subequations}
	\label{eq:}
\begin{align}
	&\mathcal{C}_{\varepsilon}\ceq \{s\in \mathbb{C}\mid\rmRe[s]>\varepsilon\},\
	\varepsilon\in(0,1/2)\\
	&\mathcal{D}_{\varepsilon,\delta}\ceq\{\delta+jr\mid-R_{\varepsilon,\delta}\leq r\leq R_{\varepsilon,\delta}\}\cup \mathcal{A}_{\varepsilon,\delta}\\
	&\mathcal{A}_{\varepsilon,\delta}\ceq\{\delta+R_{\varepsilon,\delta}e^{j\theta}\mid\theta\in \Theta\}\\
	&\Theta \ceq (-\pi/2,\pi/2),\
	R_{\varepsilon,\delta}\ceq 2\omega_\mathrm{b}\omega_\mathrm{c} L\phi_\varepsilon/(\delta-\varepsilon)\\
	&\phi_\varepsilon\ceq1/(1-\max_{\rmRe[s]=\varepsilon}|\Phi(s)|)\in [1,\infty)\\
	\label{eq:app3:def:delta}
	&\delta \in \Delta_\varepsilon\ceq\{\delta\in(\varepsilon, 1/2] \mid R_{\varepsilon,\delta}>2(\delta+\omega_\mathrm{b})\}
\end{align}
\end{subequations}
using the linear-phase low-pass filter $\Phi(s)$ in \eqref{def:CT:Phi}.
On the region $\mathcal{A}_{\varepsilon,\delta}\subset\mathcal{S}_\delta$, the non-constant filter $\Phi(s)$ satisfies
\begin{subequations}
	\label{eq:}
\begin{align}
	&|\Phi(s)|<\max_{\rmRe[s]=\delta}|\Phi(s)|<\max_{\rmRe[s]=\varepsilon}|\Phi(s)|<1\\
	&{1}/(1-|\Phi(s)|)<\phi_\varepsilon<\infty
\end{align}
\end{subequations}
based on Lemma~\ref{lem:Phi}.
Given $\varepsilon$, $\phi_\varepsilon$ is a finite constant independent of $\delta$.
Using the following relationships
\begin{subequations}
	\label{eq:app3:Gamma:elements:bounds}
\begin{align}
	\label{eq:app3:Gamma:elements:bounds:a}
	&\left|\frac{1+\Phi(s)}{1-\Phi(s)}\right|
	\leq 2\phi_\varepsilon,\ \forall s\in \mathcal{A}_{\varepsilon,\delta}.\\
	& R_{\varepsilon,\delta}>2(\delta+\omega_\mathrm{b}) \Rightarrow - (\delta+\omega_\mathrm{b}) > -R_{\varepsilon,\delta}/2\notag\\
	&\Rightarrow 1/(R_{\varepsilon,\delta} - \delta - \omega_\mathrm{b}) < 2/R_{\varepsilon,\delta},
\end{align}
\end{subequations}
the open-loop transfer function \eqref{eq:CT:Gamma} is bounded as
\begin{align}
	\label{eq:app3:absgam:upper}
	\sup_{s\in\mathcal{A}_{\varepsilon,\delta}}|\Gamma(s)|
	&\leq\frac{\omega_\mathrm{b}\omega_\mathrm{c} L\phi_\varepsilon}{R_{\varepsilon,\delta}-\delta-\omega_\mathrm{b}}
	< \frac{2\omega_\mathrm{b}\omega_\mathrm{c} L\phi_\varepsilon}{R_{\varepsilon,\delta}}
	= \delta-\varepsilon\notag\\
	& < 1/2.
\end{align}
Hence,
\begin{align}
	\label{eq:1gam:series}
	\log(1+\Gamma(s))=\Gamma(s) &+ \sum_{n=2}^\infty (-1)^{n+1} \frac{\Gamma(s)^n}{n}\\
	|\log(1+\Gamma(s))-\Gamma(s)|
	&\leq \sum_{n=0}^\infty|\Gamma(s)|^n-1-|\Gamma(s)|\notag\\
	&\leq \frac{|\Gamma(s)|^2}{1-|\Gamma(s)|}
	< 2 |\Gamma(s)|^2.
\end{align}
These and the estimation lemma give
\begin{align}
	\label{eq:}
	\left|\int_{\mathcal{A}_{\varepsilon,\delta}}\log(1+\Gamma(s))-\Gamma(s)ds\right|
	< 2\pi R_{\varepsilon,\delta} \sup_{s\in\mathcal{A}_{\varepsilon,\delta}}|\Gamma(s)|^2
\end{align}
with the length $\pi R_{\varepsilon,\delta}$ of $\mathcal{A}_{\varepsilon,\delta}$.
The limit of the upper bound equals zero as
\begin{align}
	\lim_{\delta\to\varepsilon^+}&R_{\varepsilon,\delta} \sup_{s\in\mathcal{A}_{\varepsilon,\delta}}|\Gamma(s)|^2
	\leq\lim_{\delta\to\varepsilon^+}R_{\varepsilon,\delta}\left(\frac{2\omega_\mathrm{b}\omega_\mathrm{c} L\phi_\varepsilon}{R_{\varepsilon,\delta}}\right)^2\notag\\
	&\leq \lim_{\delta\to\varepsilon^+}2\omega_\mathrm{b}\omega_\mathrm{c} L\phi_\varepsilon(\delta-\varepsilon)=0.
\end{align}
Thus, $\lim_{\delta\to\varepsilon^+}|\int_{\mathcal{A}_{\varepsilon,\delta}}\log(1+\Gamma(s))-\Gamma(s)ds|= 0$, and
\begin{align}
	\label{eq:app3:log1gam:gam:exchange}
	\lim_{\delta\to\varepsilon^+}\int_{\mathcal{A}_{\varepsilon,\delta}}\log(1+\Gamma(s))ds=\lim_{\delta\to\varepsilon^+}\int_{\mathcal{A}_{\varepsilon,\delta}}\Gamma(s)ds.
\end{align}
In $\mathbb{C}_+$, the sensitivity function $S(s)=1/(1+\Gamma(s))$ has no zeros owing to Lemma~\ref{lem:OLTF} and no poles because the QDOB is stable according to \cite{Muramatsu2025TCST} with \eqref{eq:app2:absPhi:RHP}.
Hence, $S(s)$ is analytic in $\mathbb{C}_+$.
For the analytic function $S(s)$ that has no zeros in the simply connected region $\mathcal{C}_{\varepsilon}$, there exists a single-valued analytic branch of the logarithm of $S(s)$ in $\mathcal{D}_{\varepsilon,\delta}$ according to \cite{ahlfors1979complex}.
Let $\log(S(s))$ denote this single-valued analytic branch such that $\log(S(s))=\ln|S(s)|+j\angle S(s)$.
We apply the Cauchy integral theorem to the branch $\log(S(s))$ on $\mathcal{D}_{\varepsilon,\delta}$ traversed counterclockwise:
\begin{align}
	\label{eq:app3:CauchyIntegral}
	\oint_{\mathcal{D}_{\varepsilon,\delta}}\log(S(s))ds&=0.
\end{align}
Then, \eqref{eq:app3:CauchyIntegral} can be transformed as follows
\begin{subequations}
	\label{eq:}
\begin{align}
	&\int_{\delta+jR_{\varepsilon,\delta}}^{\delta-jR_{\varepsilon,\delta}}\log (S(s))ds + \int_{\mathcal{A}_{\varepsilon,\delta}}\log(S(s))ds=0\\
	&j\int_{-R_{\varepsilon,\delta}}^{R_{\varepsilon,\delta}}\log (S(\delta+j\omega))d\omega = \int_{\mathcal{A}_{\varepsilon,\delta}}\log(S(s))ds\\
	&\int_{-R_{\varepsilon,\delta}}^{R_{\varepsilon,\delta}}\ln|S(\delta+j\omega)|d\omega = \mathrm{Im}\left[\int_{\mathcal{A}_{\varepsilon,\delta}}\log(S(s))ds\right].
\end{align}
\end{subequations}
The Bode-like sensitivity integral is calculated as
\begin{align}
	\label{eq:app3:Bode:integral}
	&\lim_{\varepsilon\to0^+}\int_{0}^{\infty}\ln|S(\varepsilon+j\omega)|d\omega
	=\lim_{\delta\to0^+}\frac{1}{2}\int_{-\infty}^{\infty}\ln|S(\delta+j\omega)|d\omega\notag\\
	&=\frac{1}{2}\lim_{\varepsilon\to0^+}\lim_{\delta\to\varepsilon^+}\int_{-R_{\varepsilon,\delta}}^{R_{\varepsilon,\delta}}\ln|S(\delta+j\omega)|d\omega\notag\\
	&=\frac{1}{2}\lim_{\varepsilon\to0^+}\lim_{\delta\to\varepsilon^+}\mathrm{Im}\left[\int_{\mathcal{A}_{\varepsilon,\delta}}\log(S(s))ds\right]\notag\\
	&=-\frac{1}{2}\lim_{\varepsilon\to0^+}\lim_{\delta\to\varepsilon^+}\mathrm{Im}\left[\int_{\mathcal{A}_{\varepsilon,\delta}}\log(1+\Gamma(s))ds\right].
\end{align}
From \eqref{eq:app3:log1gam:gam:exchange},
\begin{align}
	\label{eq:app3:Bode:1}
	&\lim_{\delta\to\varepsilon^+}\int_{\mathcal{A}_{\varepsilon,\delta}}\log(1+\Gamma(s))ds
	=\lim_{\delta\to\varepsilon^+}\int_{\mathcal{A}_{\varepsilon,\delta}}\Gamma(s)ds\notag\\
	&=j\lim_{\delta\to\varepsilon^+}\int_{-\pi/2}^{\pi/2}\Gamma(\delta+R_{\varepsilon,\delta}e^{j\theta})R_{\varepsilon,\delta}e^{j\theta} d\theta.
\end{align}
The limit and integral can be interchanged on the basis of the dominated convergence theorem as
\begin{align}
	\label{eq:app3:Bode:2}
	&\lim_{\delta\to\varepsilon^+}\int_{-\pi/2}^{\pi/2}\Gamma(\delta+R_{\varepsilon,\delta}e^{j\theta})R_{\varepsilon,\delta}e^{j\theta} d\theta\notag\\
	&=\int_{-\pi/2}^{\pi/2}\lim_{\delta\to\varepsilon^+}\Gamma(\delta+R_{\varepsilon,\delta}e^{j\theta})R_{\varepsilon,\delta}e^{j\theta} d\theta\notag\\
	&=\frac{\omega_\mathrm{b}\omega_\mathrm{c}L}{2}\int_{-\pi/2}^{\pi/2}d\theta
	=\frac{\omega_\mathrm{b}\omega_\mathrm{c}L\pi}{2},
\end{align}
which is based on
\begin{subequations}
	\label{eq:}
\begin{align}
	&\lim_{\delta\to\varepsilon^+}R_{\varepsilon,\delta}=\infty\\
	&\lim_{\delta\to\varepsilon^+}|\Phi(\delta+R_{\varepsilon,\delta}e^{j\theta})|\notag\\
	&\hspace{1em} \leq\lim_{\delta\to\varepsilon^+}\prod_{i=1}^{l}\sum_{n=-N}^{N} \left|\frac{\alpha_i(n)}{\gamma_i}\right| e^{((n-N)\bar{U}_i-\kappa)TR_{\varepsilon,\delta}\cos(\theta)}\notag\\
	&\hspace{1em} =0\\
	&\lim_{\delta\to\varepsilon^+}\Phi(\delta+R_{\varepsilon,\delta}e^{j\theta})=0,\ \forall\theta\in\Theta\\
	&\lim_{\delta\to\varepsilon^+}\Gamma(\delta+R_{\varepsilon,\delta}e^{j\theta})R_{\varepsilon,\delta}e^{j\theta}\notag\\
	&=\lim_{\delta\to\varepsilon^+}\frac{\omega_\mathrm{b} \omega_\mathrm{c} L}{2}\frac{1+\Phi(\delta+R_{\varepsilon,\delta}e^{j\theta})}{1-\Phi(\delta+R_{\varepsilon,\delta}e^{j\theta})}\frac{R_{\varepsilon,\delta}e^{j\theta}}{\delta+R_{\varepsilon,\delta}e^{j\theta}+\omega_\mathrm{b}}\notag\\
	&={\omega_\mathrm{b} \omega_\mathrm{c} L}/{2},
\end{align}
\end{subequations}
and
\begin{align}
	&|\Gamma(\delta+R_{\varepsilon,\delta}e^{j\theta})R_{\varepsilon,\delta}e^{j\theta}|
	\leq 2\omega_\mathrm{b}\omega_\mathrm{c}L\phi_\varepsilon,
\end{align}
where this upper bound is an integrable function independent of $\delta$.
From \eqref{eq:app3:Bode:integral}, \eqref{eq:app3:Bode:1}, and \eqref{eq:app3:Bode:2}, the Bode-like sensitivity integral for the QDOB is obtained as
\begin{align}
	\label{eq:}
	\lim_{\varepsilon\to0^+}\int_{0}^{\infty}\ln|S(\varepsilon+j\omega)|d\omega=-\frac{\omega_\mathrm{b}\omega_\mathrm{c} L\pi}{4}.
\end{align}

\section{Proof of Theorem~\getrefnumber{thm:Bode:Integral:QDOB:DT}}\label{sec:app:Bode:DT}
Let a simply connected region $\mathcal{C}_{\varepsilon}\subset\mathbb{C}$ be
\begin{align}
	&\mathcal{C}_{\varepsilon}=\{\tilde{z}\in \mathbb{C}\mid|\tilde{z}|<\varepsilon^{-1}<1\},\
	\tilde{z}\ceq z^{-1}.
\end{align}
Suppose the linear-phase low-pass filter $\tilde{\Phi}(z)$ in \eqref{eq:DT:Phi}, open-loop transfer function $\tilde{\Gamma}(z)$ in \eqref{eq:DT:Gamma}, and sensitivity function $\tilde{S}(z)$.
We substitute $z=\tilde{z}^{-1}$ for the functions, as follows
\begin{subequations}
	\label{eq:}
\begin{align}
	\hat{\Phi}(\tilde{z})&\ceq\tilde{\Phi}(\tilde{z}^{-1})=p_1\tilde{z}+p_2\tilde{z}^2+\cdots +p_{\bar{L}}\tilde{z}^{\bar{L}}\\
	\hat{\Gamma}(\tilde{z})&\ceq\tilde{\Gamma}(\tilde{z}^{-1})=\frac{\omega_\mathrm{c} L}{2}\frac{1+\hat{\Phi}(\tilde{z})}{1-\hat{\Phi}(\tilde{z})}\frac{\omega_\mathrm{b}T}{1+\omega_\mathrm{b}T-\tilde{z}}\\
	\hat{S}(\tilde{z})&\ceq \tilde{S}(\tilde{z}^{-1})=1/(1+\hat{\Gamma}(\tilde{z})),
\end{align}
\end{subequations}
where $p_1,\ p_2,\ \ldots \in \mathbb{R}$.
According to \cite{ahlfors1979complex}, for the analytic function $\hat{S}(\tilde{z})$ that has no zeros in the simply connected region $\mathcal{C}_{\varepsilon}$, there exists a single-valued analytic branch of the logarithm of $\hat{S}(\tilde{z})$ in $\mathcal{C}_{\varepsilon}$.
Let $\log(\hat{S}(\tilde{z}))$ denote this single-valued analytic branch such that $\log(\hat{S}(\tilde{z}))=\ln|\hat{S}(\tilde{z})|+j\angle \tilde{S}(z)$.
In the open disk $\mathcal{C}_{\varepsilon}$, the analytic branch $\log(\hat{S}(\tilde{z}))$ has the following Maclaurin series
\begin{align}
	\label{eq:}
	\log(\hat{S}(\tilde{z}))&=\log(\hat{S}(0)) + a_1\tilde{z} + a_2\tilde{z}^2+\cdots
\end{align}
with $a_1,\ a_2,\ \ldots \in \mathbb{C}$.
Accordingly,
\begin{align}
	\label{eq:}
	\log(\tilde{S}(z))=\log(\hat{S}(0)) + a_1z^{-1} + a_2z^{-2}+\cdots
\end{align}
Using this series, the contour integral of $\log(\tilde{S}(z))z^{-1}$ along $|z|=\varepsilon$ is
\begin{align}
	\label{eq::app:DT:oint:1}
	\lim_{\varepsilon\to1^+}&\oint_{|z|=\varepsilon}\log(\tilde{S}(z))z^{-1}dz\notag\\
	&=\lim_{\varepsilon\to1^+}\oint_{|z|=\varepsilon}\log(\hat{S}(0))z^{-1} + a_1z^{-2} + \cdots  dz\notag\\
	&=j\lim_{\varepsilon\to1^+}\int_{0}^{2\pi}\log(\hat{S}(0)) + a_1\varepsilon^{-1}e^{-j\Omega}+\cdots d\Omega\notag\\
	&=j2\pi\log(\hat{S}(0)).
\end{align}
Meanwhile, the contour integral becomes
\begin{align}
	\label{eq:app:DT:oint:2}
	\lim_{\varepsilon\to1^+}&\oint_{|z|=\varepsilon}\log(\tilde{S}(z))z^{-1}dz\notag\\
	&=\lim_{\varepsilon\to1^+}\int_{0}^{2\pi}j\ln|\tilde{S}(\varepsilon e^{j\Omega})|-\angle \tilde{S}(\varepsilon e^{j\Omega})d\Omega.
\end{align}
From the imaginary parts of \eqref{eq::app:DT:oint:1} and \eqref{eq:app:DT:oint:2}, the Bode-like sensitivity integral for the discrete-time representation is
\begin{align}
	\label{eq:}
	&\lim_{\varepsilon\to1^+}\int_{0}^{2\pi}\ln|\tilde{S}(\varepsilon e^{j\Omega})|d\Omega
	=2\pi\rmRe[\log(\hat{S}(0))]\notag\\
	&=2\pi\ln(2+2\omega_\mathrm{b}T)
	-2\pi\ln(2+2\omega_\mathrm{b}T+\omega_\mathrm{b}\omega_\mathrm{c} LT).
\end{align}


\end{document}